\documentclass{ws-ijmpd}
\usepackage[super,compress]{cite}

\usepackage{graphicx}	

\usepackage{amsthm,mathtools}
\usepackage{amsmath}	

\usepackage{url}
\usepackage{hyperref}

\newcommand{\gttup}{g^{tt}}
\newcommand{\gtpup}{g^{t\phi}}
\newcommand{\gppup}{g^{\phi \phi}}
\newcommand{\gtt}{g_{tt}}
\newcommand{\gtp}{g_{t\phi}}
\newcommand{\gpp}{g_{\phi \phi}}
\newcommand{\utup}{U^t}
\newcommand{\upup}{U^\phi}
\newcommand{\ut}{U_t}

\newcommand{\at}{A_t}
\newcommand{\aphi}{A_\phi}

\newcommand{\Tab}{T^{\alpha \beta}}
\newcommand{\Tabmat}{T^{\alpha \beta}_{\text{MAT}}}
\newcommand{\Tabem}{T^{\alpha \beta}_{\text{EM}}}
\newcommand{\Fab}{F^{\alpha \beta}}
\newcommand{\Fabext}{F^{\alpha \beta}_{\text{EXT}}}
\newcommand{\Fabint}{F^{\alpha \beta}_{\text{INT}}}
\newcommand{\gab}{g^{\alpha \beta}}

\newcommand{\At}{A_{t}}
\newcommand{\Ap}{A_{\phi}}

\begin{document}

\markboth{A. Trova}
{Equilibrium of charged fluid around a Kerr black hole immersed in a magnetic field: variation of angular momentum}

%
\catchline{}{}{}{}{}
%

\title{Equilibrium of charged fluid around a Kerr black hole immersed in a magnetic field: variation of angular momentum}

\author{Audrey Trova,\thanks{E-mail: audrey.trova@zarm.uni-bremen.de}}

\address{University of Bremen,  Center of Applied Space Technology and Microgravity (ZARM), Am Fallturm 2\\
28209 Bremen,
Germany\\
audrey.trova@zarm.uni-bremen.de}

\maketitle

\begin{history}
\end{history}

\begin{abstract}
The present work presents analytically constructed equilibrium structures of charged perfect fluids orbiting Kerr Black holes embedded in an asymptotically uniform magnetic field. Our focus is on the effect of the non-constant angular momentum distribution through the disk, as well as its combined effect with the external magnetic field and the fluid charge. We demonstrate that the three parameters of our study have a significant impact on the various features of the accretion disk: the shape, the size of the disk and the characteristic of the fluid, as the pressure and the rest-mass density. Through our investigation, we observe substantial deviations from both the uncharged thick disk model and the charged disk model with constant angular momentum.
\end{abstract}


\ccode{PACS numbers:}


\section{Introduction}	

AGN (Active Galactic Nuclei), quasars, and X-ray binaries are some of the most compelling and enigmatic sources of energy in the Universe.  They are powered by accretion processes between them and the accretion disk orbiting around them. These phenomena involve the infall of matter onto a central black hole or a compact object, resulting in the release of vast amounts of energy in the form of radiation and high-speed jets. To gain a comprehensive understanding of these highly energetic systems, the investigation of the accretion disks becomes crucial. The theoretical description of accretion disks (ADs) and their dynamic behavior begins with the specification of the underlying matter model. The choice of the matter model is crucial as it dictates the physical properties, behavior, and interactions of the matter within the disk. Such astrophysical objects are commonly described using fluid approaches.\cite{Punsly08,PaLeReRe09,ReZa13} The corresponding matter model has to be embedded into a space-time geometry which in most cases is the gravitational field of the compact object. In real astrophysical system, the black-hole accretion disk system is not isolated but is surrounded by electromagnetic fields and radiation. It is well established that the effects of the magnetic field have a crucial impact on the accretion disks. A global magnetic field seems to be essential in various mechanisms of jets and/or wind. 

In the universe, the overall matter is indeed electrically neutral. However, in astrophysical environments one frequently encounters ionized plasma, consisting of both positive ions and negative electrons. This ionization can occur due to various processes, such as high temperatures, intense radiation, or interactions with nearby stars. The presence of magnetic fields, either of galactic origin or from nearby celestial bodies, plays a crucial role in such environments. The magnetic fields can induce the separation of charges within the plasma, causing lighter charged particles to be more strongly influenced by the magnetic forces and thereby accelerating them. As a result, they may gain sufficient energy to escape the region, while the heavier charged particles remain in the vicinity. This separation of charges can lead to the formation of what is known as a "charged fluid." Noteworthy examples of such scenarios include the pulsar magnetosphere and black holes immersed in external magnetic fields, as studied by Ref.~\refcite{1993A&A...274..319N} and Ref.~\refcite{2002A&A...384..414P}.
In a series of paper by Ref.~\refcite{Kovar11,KovarTr14,KoSlaCreStuKaTro16,SchTrHacLam18,Trova18,2020Trova}, the equilibrium of charged thick accretion disk in an gravito-electromagnetic background has been studied. The selected system for this study comprises a black hole surrounded by an asymptotically homogeneous magnetic field, likely generated by a magnetar situated at a considerable distance. Within this arrangement, we observe a potential magnetic field strength denoted as $\tilde{B}$, with an upper limit of approximately $10^6 \mathrm{T}$ (or $B \lesssim 10^{-10}$ in dimensionless units, assuming $\tilde{M}=M_{\odot}$). This limitation remains within reasonable bounds for treating the external magnetic field as a test field as discussed in the Ref.~\refcite{Kolos_2015}.A comprehensive description of this system is available in the referenced literature, specifically in the Ref.~\refcite{KovarTr14}. It has been showed that equatorial bounded fluid structures, typically used as initial condition in GRMHD simulation, exist when including the charge of the fluid. The thick disk model also known as Polish doughnut forms the foundation for the current approach Ref.~\refcite{AbrJaSi78,AbraFra2013,ReZa13}, and the subsequent papers have built upon and extended its concepts to formulate the specific model being studied in this paper. The CAD model reproduced the same features as the one possesses by the thick accretion model: (i) a maximum of pressure relevant for epicyclic oscillatory modes. These oscillations can be associated with the motion of fluid elements in the disk, and their presence can have implications for various phenomena, including quasi-periodic oscillations and instabilities in the accretion process. (ii) A funnel-like structure along the rotational axis. This funnel region is particularly important as it is relevant for the collimation and acceleration of powerful jets observed in astrophysical objects like AGN and quasars. The magnetic fields in this region can play a crucial role in guiding and channeling the outflowing matter into highly collimated and energetic jets. (iii) A cusp corresponding to the self-crossing point of one equipotential surface relevant for inner
boundary conditions.

The CAD model involves the addition of a global non-zero charge to the circulating fluid within the thick disk. As a result, the charged fluid interacts with the external background electromagnetic field originating from an external source or with the one coupled to the central object. The other main characteristic of this model is the assumption of zero conductivity. The electromagnetic field is then connected with the motion of the fluid. The charged particle is sticking to the fluid matter. This hypothesis is the opposite of the usual force-free approximation of the ideal magnetohydrodynamic (MHD) commonly used in GRMHD simulation. The ideal MHD is indeed an excellent and reasonable approximation in many astrophysically relevant situations involving fluids in motion. However, in catastrophic events as a merger of two neutron stars or neutron star-black hole, plasma with high temperature and low density can be produced; thus in such a regime, non-ideal effects can appear \cite{10.1111/j.1365-2966.2009.14454.x}. 
Moreover, one can envisage a pertinent scenario, such as dusty plasmas, wherein a neutral fluid accommodates a limited number of free charges. When subjected to an external electromagnetic field, these charges undergo movement, thereby generating a current. However, in cases where the fluid attains sufficient density, the mobility of the charges diminishes, potentially leading to an almost negligible conductivity \cite{REI65}. In that cases, a model of non-conductive charged fluid is relevant and are of some interest as initial conditions for numerical simulations. Recently, GRMHD has been enlarged to include non-zero resistive fluids, that is, to non-idal GRMHD, and has been added to the BHAC (Black Hole Accretion Code), a multidimensional GRMHD module for the MPI-AMRVAC (Adaptive Mesh Refinement - Versatile Advection Code) framework \cite{2019ApJS..244...10R}. It is known that resistivity plays an important role in the magnetic reconnection and plasmoid growth in BH AD-jets systems. The magnetic field can change its topology through magnetic reconnection. This results in the dissipation of the released magnetic energy, which accelerates particles, causing non-thermal emission. Those emissions are present in different sources, as in Sgr A*, the BH at the center of the Milky Way. This non-thermal emission is one of the uncertainties of the ideal GRMHD models of the EHT observations of the accretion disk of M87 and the supermassive BH at the center of the galaxy M87. \cite{2019ApJS..244...10R} showed that a high resistivity affect the evolution of the system by diminishing the MRI-induced turbulence. Those results where consistent with the work by Ref.~\refcite{2017ApJ...834...29Q}. In that situation, the infinite resistivity solutions can be interesting as initial conditions to those simulations.

Prior works have built equilibrium solutions using constant angular velocity or constant angular momentum. Based on the output of GRMHD simulation Ref.~\refcite{Machida_2008,Fragile_2009}, the angular distribution in the disk is not constant through the disk. The purpose of this paper is to develop equilibrium solutions of charged perfect fluids with non-constant angular momentum profiles and to analyze how it affects the features of accretion disks. The Section \ref{sec:level2} establishes the general equations and assumptions of the CAD model. The equilibrium solution modelling the accretion disk can be found in Sec. \ref{sec:EqSol}. The Sec. \ref{sec:Pressure} describes the fluid’s properties and how they are affected by the magnetic field, the charge of the fluid and the angular momentum distribution. In this work, we use geometrized unit system $G=1=c$ and metric signature $(-,+,+,+)$.




\section{\label{sec:level2} Description of the CAD model}
\subsection{\label{sec:level21} General equations}
A matter model typically involves defining the fluid's properties, such as density, pressure, temperature, and velocity distribution. In many cases, accretion disks are modeled as perfect fluids, where the fluid is assumed to be continuous, with no dissipation or viscosity effects. This is a common and simple approach that can provide valuable insights into the overall behavior of the disk. In this work the accretion disk is modeled by a charged perfect fluid. Thus the shape of the disk is described by the conversation law and the Maxwell equations.
\begin{align}
&\nabla_{\beta} \Tab =0 \label{eq:Continuity1}\,, \\
&\nabla_{\beta} \Fab =4 \pi J^{\alpha} \label{eq:Maxwell}\,, \\
&\nabla_{\left(\gamma\right.}F_{\left.\zeta\nu\right)}=0 \label{eq:Property} \,.
\end{align}
The stress-energy tensor $\Tab$ combines the energy-momentum contributions from both matter, $\Tabmat$, and electromagnetic fields $\Tabem$. They are expressed as follows:
\begin{align}
&\Tabmat = (\epsilon + p)U^{\alpha}U^{\beta}+p\gab\,, \\
&\Tabem = \frac{1}{4\pi}\left(F^{\alpha}_{\;\;\gamma}F^{\beta \gamma}-\frac{1}{4}F_{\gamma \delta}F^{\gamma \delta}\gab\right)\,.
\end{align}
$p$ is the pressure of the fluid and $\epsilon$ the energy density. $U^{\alpha}$ is the four-velocity vector. The electromagnetic tensor $\Fab$, as well as the energy momentum tensor can be split into two terms.
\begin{itemize}
    \item The tensor describing the external electromagnetic forces acting on the accretion disk, $\Fabext$.
    \item And the tensor describing the self-magnetic interactions due to the charged particles of the disk, $\Fabint$.
\end{itemize}
We neglect the self-electromagnetic interactions as a first step, thus we have $\Fabint<<\Fabext$. That means that the charge and the magnetic field produced by the disk itself have no impact on the disk and on the spacetime.
The electromagnetic field tensor $\Fabext$ can be expressed in terms of its electromagnetic potential vector $A_\nu$ as follows:
\begin{equation}
\label{eq:ElectroMagTensor}
\Fabext = g^{\alpha\zeta} g^{\beta \nu}(\nabla_{\zeta}A_{\nu}-\nabla_{\nu}A_{\zeta})\,.
\end{equation}
 The four-current density vector $J^{\alpha}$ depends on the charged density $\rho_q$, on the electrical conductivity $\sigma$ and on the four-velocity $U^{\alpha}$. Its expression is given through the Ohm's law.

\begin{equation}
 \mathbf{J^{\alpha}=\rho_qU^{\alpha}+\sigma \Fab U_{\beta}\,.}
\end{equation}
Using Eqs. (\ref{eq:Continuity1}), (\ref{eq:Maxwell}), (\ref{eq:Property}), the stress-energy tensor decomposition and the assumptions mentioned above, we get:
\begin{align}
 \nabla_\nu \Tabmat=\Fabext J_\nu \, .
 \label{masterformular}
 \end{align}
Please note that the above equations assume the use of natural units, where the speed of light, $c$, and the gravitational constant, $G$, are set to unity ($c = G = 1$). 
\subsection{\label{sec:level22} General assumptions of the CAD model}
The charged fluid is assumed to exhibit the following characteristics: it is stationary, axially symmetric and in circular rotation around a charged black hole, while embedded in a uniform magnetic field. The circular motion of the fluid leads to a specific form for the four-velocity vector, given by $U^\alpha=(\utup,0,0,\upup)$. 
The axially symmetric and stationary nature of the system implies that $\partial_t=\partial_{\phi}=0$. To satisfy this requirement, one of the options is to set the conductivity $\sigma$ of the fluid to zero. This assumption contrasts with the ideal magnetohydrodynamic (MHD) model, where conductivity is infinite. In this model, the charged particles are assumed to be sticking to the matter and following the fluid motion. We have to note that, in case of a non zero conductivity radial currents should arise. This is not consistent with the assumption of circular motion in thick disk model. The charged fluid is treated as a test-fluid, implying that it does not significantly influence the spacetime. Additionally, the external magnetic field has a negligible effect on the spacetime metric. Finally, the electromagnetic field, considered here, has to be axially symmetric and stationary, leading to a specific form for the electromagnetic vector potential, given by $A_{\zeta}=(A_t,A_\phi,0,0)$. Using our assumption the set of partial equations can be rewritten as:
\begin{equation}
\begin{aligned}
    &\partial_r W = \partial_r \ln \utup - \frac{\ell \partial_r \Omega}{1-\Omega \ell} +\frac{\rho_q \utup}{p+\epsilon} \left(\partial_r \at+\Omega \partial_r \aphi \right), \\ 
    &\partial_\theta W = \partial_r \ln \utup - \frac{\ell \partial_r \Omega}{1-\Omega \ell} +\frac{\rho_q\utup}{p+\epsilon} \left( \partial_\theta \at+ \Omega \partial_\theta \aphi\right) \, \\
    \label{eq:PartialDiff}
\end{aligned}
\end{equation}
with 
\begin{equation}
\mathbf{
    \partial_{\alpha} W=\frac{\partial_{\alpha} p}{p+\epsilon}, \quad \text{with} \quad \alpha\in \left\{{r,t}\right\},}
    \label{eq:WtoRho}
\end{equation}
where $W$ is the effective potential. When assuming no magnetic field, the last term of each partial derivative disappear and the polish doughnut description can be found.
The specific angular momentum and angular velocity can be defined in terms of the four velocity components as
\begin{align}
\ell=-\frac{U_\phi}{U_t},\quad \quad \Omega=\frac{U^\phi}{U^t}.
\label{lomega}
\end{align}
They are connected with each other by the equation
\begin{align}
\Omega=-\frac{\ell\,g_{tt}+\gtp}{\ell\,\gtp+\gpp}\,.
\end{align}

\subsection{Equipotential surfaces}
As Ref.~\refcite{Jaroszynski1980} shown the uncharged part of the partial derivatives can be rewritten in terms of the derivative of the metric. It follows that, in the CAD model, we can achieve the same transformation:
\begin{equation}
    \begin{aligned}
        &\partial_r W = \frac{\partial_r \gttup-2\ell \partial_r \gtpup+\ell^2\partial_r \gppup +q(r,\theta)f \left(\partial_r \at+\Omega \partial_r \aphi \right)}{2(1-\Omega\ell)}\\ 
        &\partial_\theta W = \frac{\partial_\theta \gttup-2\ell \partial_\theta \gtpup+\ell^2\partial_\theta \gppup +q(r,\theta)f \left(\partial_\theta \at+\Omega \partial_\theta \aphi \right)}{2(1-\Omega\ell)},
        \label{eq:PartialDiff2}
    \end{aligned}
\end{equation}
with $q(r,\theta)=2 \frac{\rho_q}{(p+\epsilon)\ut}$ is a function linked to the specific charge density of the CAD and $f=\gttup-\ell \gtpup$. By dividing  \ref{eq:PartialDiff2}a and \ref{eq:PartialDiff2}b, we get
\begin{equation}
    \frac{\mathrm{d}\theta}{\mathrm{d}r}=F(r,\theta),
\end{equation}
where $F(r,\theta)$ is known in a closed form, once we made the following assumptions: (i) the gravitational field (ii) the angular momentum distribution, (iii) the specific charge density $\rho_q$ and (iv) the test external magnetic field.

\subsubsection{Gravitational field $\&$ Angular momentum distribution}
\label{sec:AngMom}
In this study, the gravitational field is assumed to be generated by a Kerr black hole, described by the Kerr geometry in standard Boyer-Lindquist coordinates as follows:

\begin{equation}
ds^2 = -\left(1-\frac{2r}{\Sigma}\right) dt^2+\frac{\Sigma}{\Delta}dr^2+\Sigma d\theta^2+\left(r^2+a^2+\frac{2 r a^2}{\Sigma}\sin^2\theta\right)\sin^2\theta d\phi^2-\frac{4r a \sin^2\theta}{\Sigma}dt d\phi
\end{equation}
where $\Sigma=r^2+a^2\cos^2\theta$, $\Delta=r^2-2r+a^2$ and $a$ represents the spin parameter of the black hole. We identify the radius $r_0$ at which $\partial_rW=0$ and $\partial_r^2W=0$, and its corresponding value of angular momentum is denoted as $\ell_0$. The choice of the inner radius of the accretion disk is made within a range where bounded solutions with a cusp are allowed. In the uncharged case, this area corresponds to angular momenta values in the range $\ell_{\rm ms} > \ell > \ell_{\rm mb}$. Here, the index $\rm ms$ stands for marginally stable (where $\partial_rW=0$ and $\partial_r^2W=0$), and the index $\rm mb$ stands for marginally bounded (where $W=0$ and $\partial_rW=0$). Since we do not have access to the explicit form of $W$ in the CAD model at this stage, we approximate the inner edge of the disk to be located in the vicinity of the corresponding radius of the marginally stable orbit in the charged case, denoted here as $\ell_0$.This chosen inner edge coincides with the cusp, which is the location where accretion onto the black hole can occur. To achieve this, we use a slightly modified version of the distribution proposed by Ref.~\refcite{Qian2009}, adapted to the CAD model.
\begin{align}
\label{eq:AngMom}
 \ell(r,\theta)=
   \left\{
  \begin{array}{@{}ll@{}}
  \ell_{\mathrm{in}}\left(\sin{\theta}\right)^{2\delta}, & r<r_{\rm 0}\\
  \ell_{\mathrm{in}}\left(\frac{L(r)}{\ell_{0}}\right)^{\beta}\left(\sin{\theta}\right)^{2\delta}, & r\geq r_{\rm 0}, 
    \end{array}\right.
\end{align}
with 
\begin{align}
   0\leq \beta\leq 1, \quad -1\leq \delta \leq 1.
\end{align}
$L(r_{\mathrm{in}})=\ell_{\mathrm{in}}$ is the value of the angular momentum at the inner edge, and $L(r)$ is the angular momentum in the equatorial plane defined by the condition $\partial_r W=0$. This condition ensures that both the inner edge and the center of the disk are located at extrema of the effective potential $W$. Consequently, the inner part of the disk follows a constant angular momentum distribution, while the central region exhibits super-Keplerian motion up to the center. In the outer part of the disk, the motion becomes sub-Keplerian. These specific characteristics have been observed in various numerical simulations of accretion disks \cite{Machida_2008, Fragile_2009}. When $\beta=\delta=0$, the $\ell$-constant case is retrieved. About the parameters $\beta$ and $\delta$, it is important to note that the positions of the inner edge and the center of the disk are solely dependent on $\beta$. 

\subsubsection{Magnetic field $\&$ charge density}
Along with the gravitational field of the Kerr black hole, the CAD is embedded into an external asymptotically uniform magnetic field. This scenario  is described by Wald's test-field solution of the Maxwell equations, and it involves the following vector potential components:
\begin{align}
&\At = \frac{B}{2}\left(\gtp+2a\gtt\right) \\
&\Ap = \frac{B}{2}\left(\gpp+2a\gtp\right). 
\end{align}
Here, $B$ represents the strength of the external magnetic field, and $a$ is the spin parameter of the Kerr black hole. It is assumed that the charge of the black hole due to the influence of the magnetic field is zero \cite{Wald74,Stuchlik2016}. This assumption finds support in the case of astrophysical black holes, where various studies have demonstrated that even if there is sufficient plasma to form a charged black hole, it will rapidly discharge over a very short timescale \cite{Cardoso_2016}.

The specific charge density of the fluid, along with the overall charge density, is considered a free parameter in our model.  The only constraint imposed by our assumptions is that the magnetic field produced by the CAD should be in order of magnitude less than the external magnetic field. As starting point, to simplify the analysis, we make the assumption that the function $q(r,\theta)=k$ is constant throughout the system.  Thus the relationship between the pressure and the charge density is given by:
\begin{equation}
\mathbf{
    \rho_q=k (p+\epsilon)U_t}.
    \label{eq:ChargeP}
\end{equation}
The constant $k$ is here to quantify much the accretion disk is charged. Once we normalize the equations given in Eq. \ref{eq:PartialDiff2}, we establish a new constant $\mu=kB$, which becomes linked to both the strength of the magnetic field and the magnitude of the charge associated with the disk.


\subsection{Mass density, Pressure and Energy density}
Given the angular momentum distribution, we have obtained the equipotential surfaces which lies between the inner edge and the center of the disk. By using the expression of the equipotential in the equatorial plane, integrated from \ref{eq:PartialDiff2}a between the inner edge (index $\rm in$) and the center (index $c$),
\begin{equation}
    W_{\mathrm{Eq}}(r)=\int_{r_\mathrm{in}}^{r_\mathrm{c}}\partial_rW dr,
\end{equation}
we can build all the equipotential surfaces in all the space \cite{SolerFont18}. From the equipotential surfaces, we can obtain various characteristics of the disk by solving Eq. \ref{eq:WtoRho}. Our work is restricted to the case of adiabatic disks with no radiation, then the energy density is linked to the pressure and to the mass density by

\begin{equation}
    \epsilon=\rho + n p,
    \label{eq:EnergyP}
\end{equation}
with $n$ is  the adiabatic constant. Moreover the fluid follows an adiabatic pressure-mass density relation:
\begin{equation}
    p = K \rho^{\left(1+\frac{1}{n}\right)}
    \label{eq:PressureRho}
\end{equation}
By examining the preceding equations along with Equation \ref{eq:WtoRho}, we can deduce the expressions for mass-density, pressure, and energy density. Additionally, combining Equations \ref{eq:WtoRho}, \ref{eq:EnergyP}, and \ref{eq:PressureRho}, we establish the following correlation between the effective potential $W$ and the rest-mass density:
\begin{equation}
    \rho=\left(\frac{e^{W_{\rm in}-W}-1}{K(n+1)}\right)^n,
\end{equation}
where $W_{\rm in}$ is the value of the effective potential at the inner edge.
The present calculation yields expressions for pressure and energy density. Subsequently, utilizing Equation \ref{eq:ChargeP}, we can construct charge density profiles. Notably, the equi-potential surfaces align with equi-rest-mass density, equi-pressure, and equi-energy density surfaces. Moreover, due to $\partial_{\mu}U_t=0$, the equi-charge density surfaces also coincide. The maximal values for all physical quantities correspond precisely with the center of the disk.
\section{Equilibrium solution}
\label{sec:EqSol}

\subsection{Exploration of the space parameter of the solutions}

\subsubsection{Constant angular momentum}

\begin{figure}
    \centering
    \includegraphics[width=\hsize]{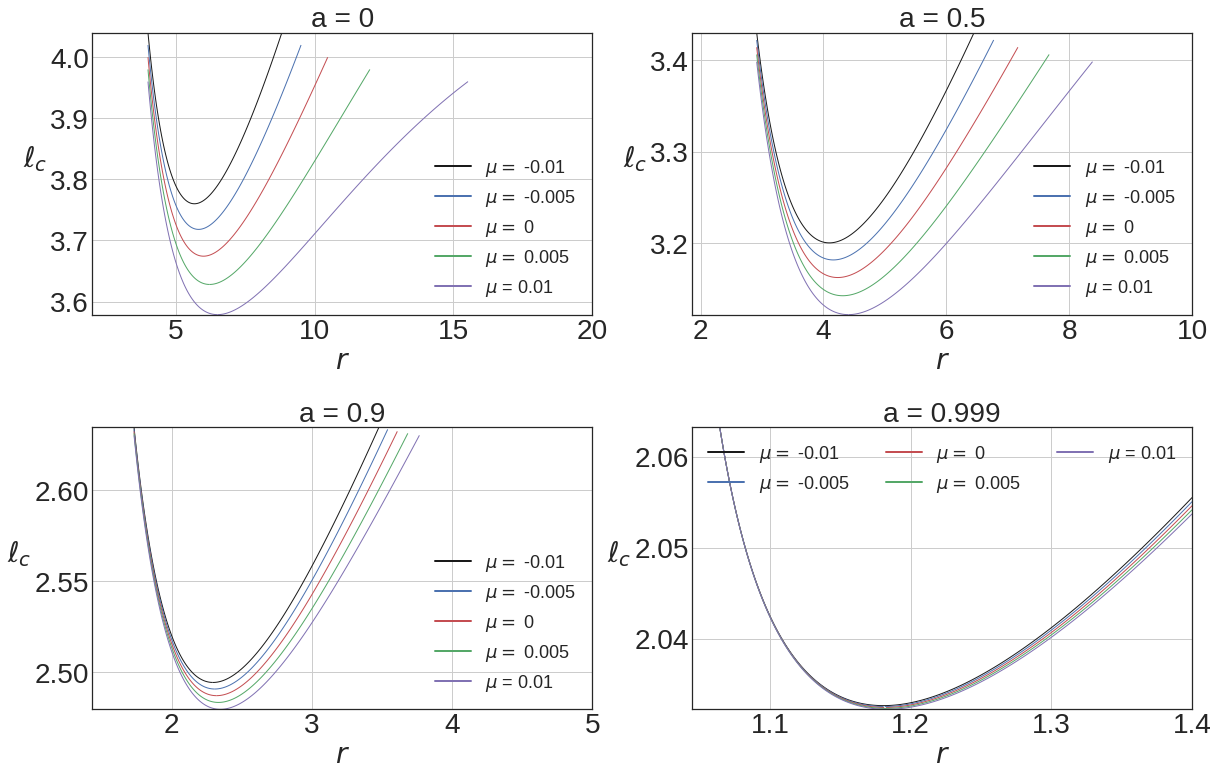}
    \caption{The curves are representing where $\partial_rW$ is zero. The minimum of each curve corresponds to the case where the disk is a ring of matter. The left branches relatively to that minimum point show the variation with the inner edge. The right branches is the variation of the center of the disk .}
    \label{fig:fig1}
\end{figure}

In this section, our primary focus is on investigating the constant angular momentum case $\ell_c,$ which has not been previously explored in this specific context. We aim to understand how the electromagnetic parameter $\mu$ influences the position of the cusp and the center of the system. The findings are presented graphically in Fig. \ref{fig:fig1}. For different values of $\mu$ and spin $a$, we plot the zeros of the first partial derivative. These zeros indicate the radial locations of the cusp (located on the left branches with respect to the curve's minimum at $(r_0,\ell_0)$) and the center (located on the right branches).
The color red in the figure corresponds to the Schwarzschild case when $a=0$. We observe that with respect to $\ell_c$, all the curves follow a similar pattern as in the Schwarzschild case. As the constant angular momentum $\ell_c$ increases, the cusp moves inward toward the black hole. Conversely, the center moves away from the central mass.
Furthermore, we analyze the effect of the parameter $\mu$ on each plot. It becomes evident that both the cusp and the center experience more pronounced effects as $\mu$ increases. However, the center is pushed away more than the cusp is moved inward. Consequently, as $\mu$ increases, the distance between these two points expands, resulting in a larger disk.
Additionally, we observe that $\ell_0$, the lower bound of the region within which bound solutions can be found, decreases as the electromagnetic parameter $\mu$ increases. Unfortunately, without access to the upper bound, we are unable to determine the behavior of the area at higher values of $\mu$.
It is worth noting an interesting observation related to the black hole's spin. The spin significantly reduces the effect of the electromagnetic parameter. As the black hole rotates faster, the difference between the curves becomes less apparent.


\subsubsection{Non-constant angular momentum}
\label{sec:EqPlaneNonConst}
\begin{figure}

\includegraphics[width=\hsize]{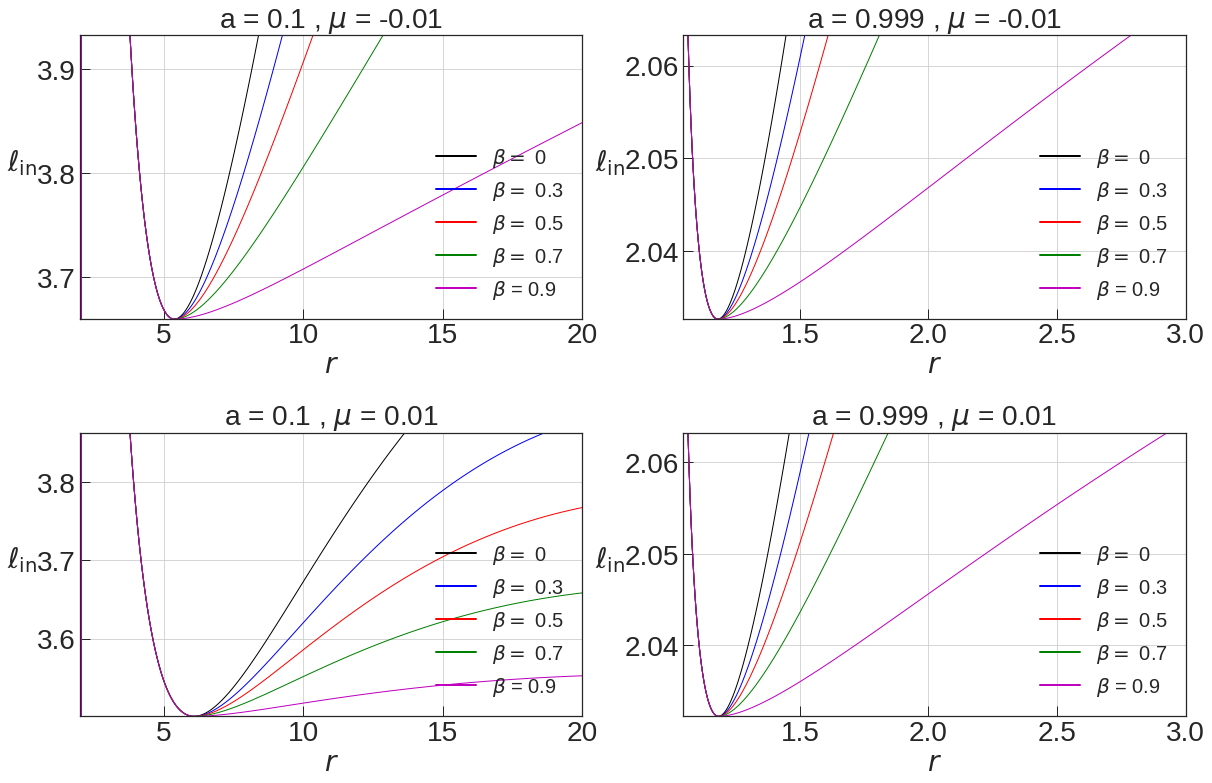}
\caption{The curves are representing where $\partial_rW$ is zero in the equatorial plane. The minimum of each curve corresponds to the case where the disk is a ring of matter. The left branches relatively to that minimum point show the variation with the inner edge. The right branches is the variation of the center of the disk.}
    \label{fig:fig2} 
\end{figure}

As mentioned in Section \ref{sec:AngMom}, our analysis is based on the angular momentum distribution given by Equation \ref{eq:AngMom}, which depends on two parameters $\beta$ and $\delta$. In this section, we will focus on examining the behavior of the inner cusp and the center, so we will solely consider the influence of the parameter $\beta$ (see in Fig. \ref{fig:fig2}). In the preceding section, it was asserted that the roots of $\partial_rW$ within the equatorial plane serve as indicators for the inner edge and center of the disk. Using $\theta=\pi/2$, the dependence with $\delta$ in the non-constant angular momentum used in this work disappears, leaving only $\beta$. The effect of $\delta$ will be noticeable only in the external part of the disk where $r \geq r_{\mathrm{0}}$.


Indeed, similar to the case with a non-rotating black hole, the effect of the electromagnetic parameter $\mu$ is attenuated. The same phenomenon applies to the influence of the parameter $\beta$ on the disk's center, which is displaced from the black hole. Consequently, as $\beta$ increases, the radial distance between the inner edge and the center also increases, resulting in a larger radial size of the disk. It is important to note that as $\beta$ increases, the angular momentum distribution approaches $L(r),$ which is constructed from $\partial_r W=0.$ This behavior leads to lower pressure gradients within the disk. As a consequence, we can anticipate the disk becoming thinner with increasing $\beta$. In the next section, we will develop the solution across the entire space, providing us with valuable insights into our expectations regarding the behavior of the disk in response to the variations in $\beta$ and other parameters.

\subsection{Solution in the entire space}


In this section, we construct the CAD throughout the entire space. The equipotential surfaces we consider are not only equi-pressure but also equi-mass-density and equi-energy density surfaces. The surface of the disk is determined at the inner edge, where $W-W_{\rm in}=0,$ which implies that both pressure, $p$, and mass density, $\rho$, are zero at this point. This condition helps us identify the extent of the disk and its boundaries.

\subsubsection{Constant angular momentum}
\label{sec:MapConst}
\begin{figure*}
    \centering
    \includegraphics[width=\hsize]{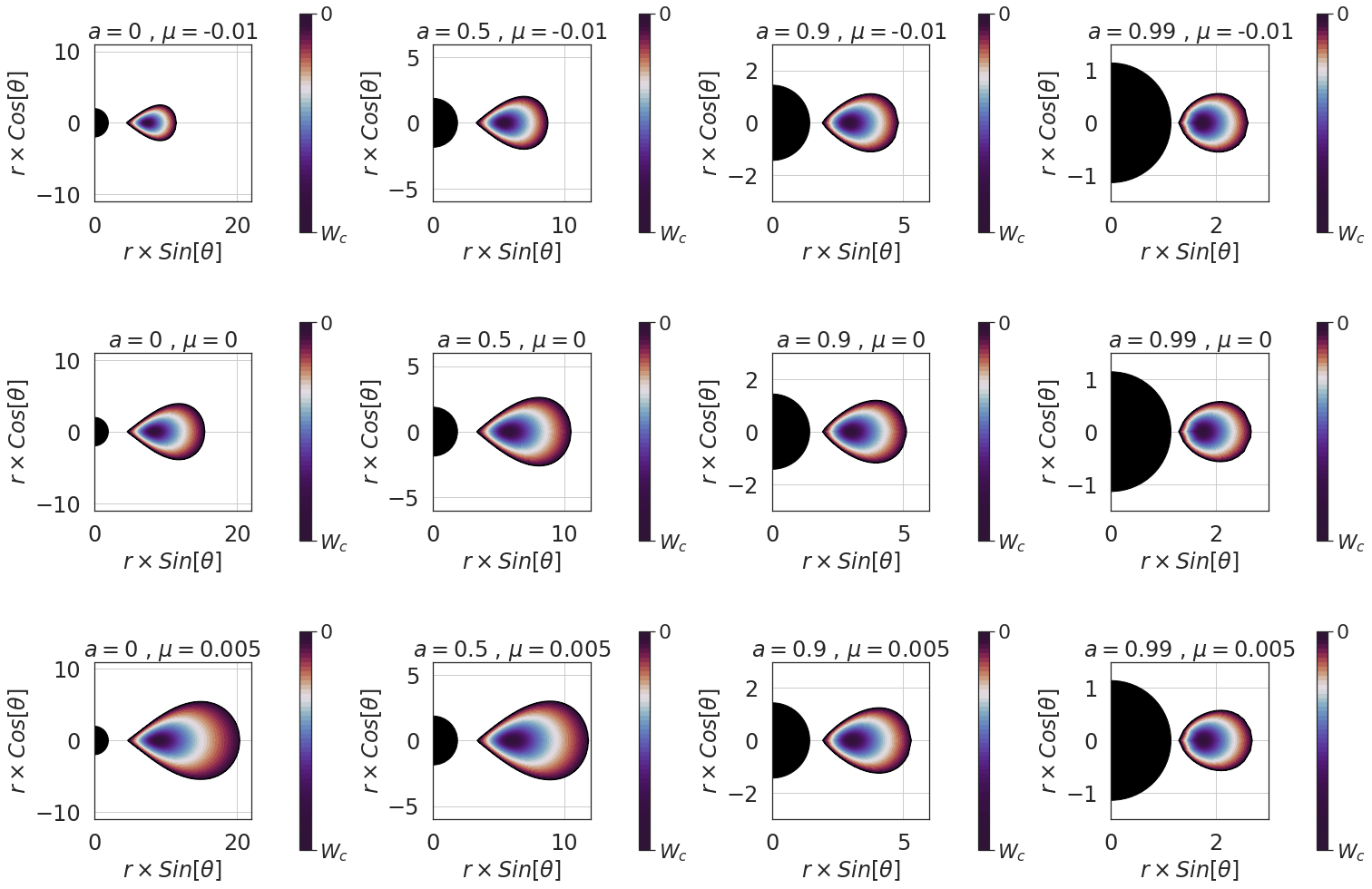}
    \caption{Map of the equipotential surfaces for various combination of the spin and the electromagnetic parameter. All along the plots the angular momentum is assumed to be constant.}
    \label{fig:EquiWConst}
\end{figure*}

Fig. \ref{fig:EquiWConst} shows the equipotential surfaces when assuming a constant angular momentum for various combinations of the spin and the electromagnetic parameter.

In the Fig. \ref{fig:EquiWConst}, the electromagnetic parameter $\mu$ increases from a negative value to a positive value from the top row to the bottom row. Additionally, the spin of the black hole increases from the left column to the right column, ranging from a Schwarzschild black hole to a fast rotating black hole. The second row displays an uncharged accretion disk for reference. Upon analyzing the Fig. \ref{fig:EquiWConst}, we can confirm that, in this particular case, the presence of the electromagnetic field significantly influences the size and shape of the accretion disk. Specifically, a negatively charged disk exhibits a smaller radial and vertical extension, while positively charged structures have a larger radial and vertical size. Moreover, the last column, representing fast rotating black holes, reveals that the equipotential surfaces appear similar regardless of the value of $\mu.$ This indicates that the spin of the black hole also mitigates the effect of the electromagnetic parameter. Furthermore, the shape of the disk is strongly affected by the black hole's spin. Increasing the spin leads to a disk that more closely resembles a torus-like structure. When the electromagnetic parameter $\mu$ is negative, the potential hole becomes wider, resulting in a more evenly distributed matter throughout the disk. Conversely, for positive values of $\mu,$ the potential hole becomes deeper, leading to a concentration of matter between the inner edge and the center of the disk. 

Overall, these observations provide valuable insights into how the combination of spin and the electromagnetic field influences the size, shape, and matter distribution within the accretion disk.

\subsubsection{Non-constant angular momentum}

\begin{figure*}
    \centering
    \includegraphics[width=\hsize]{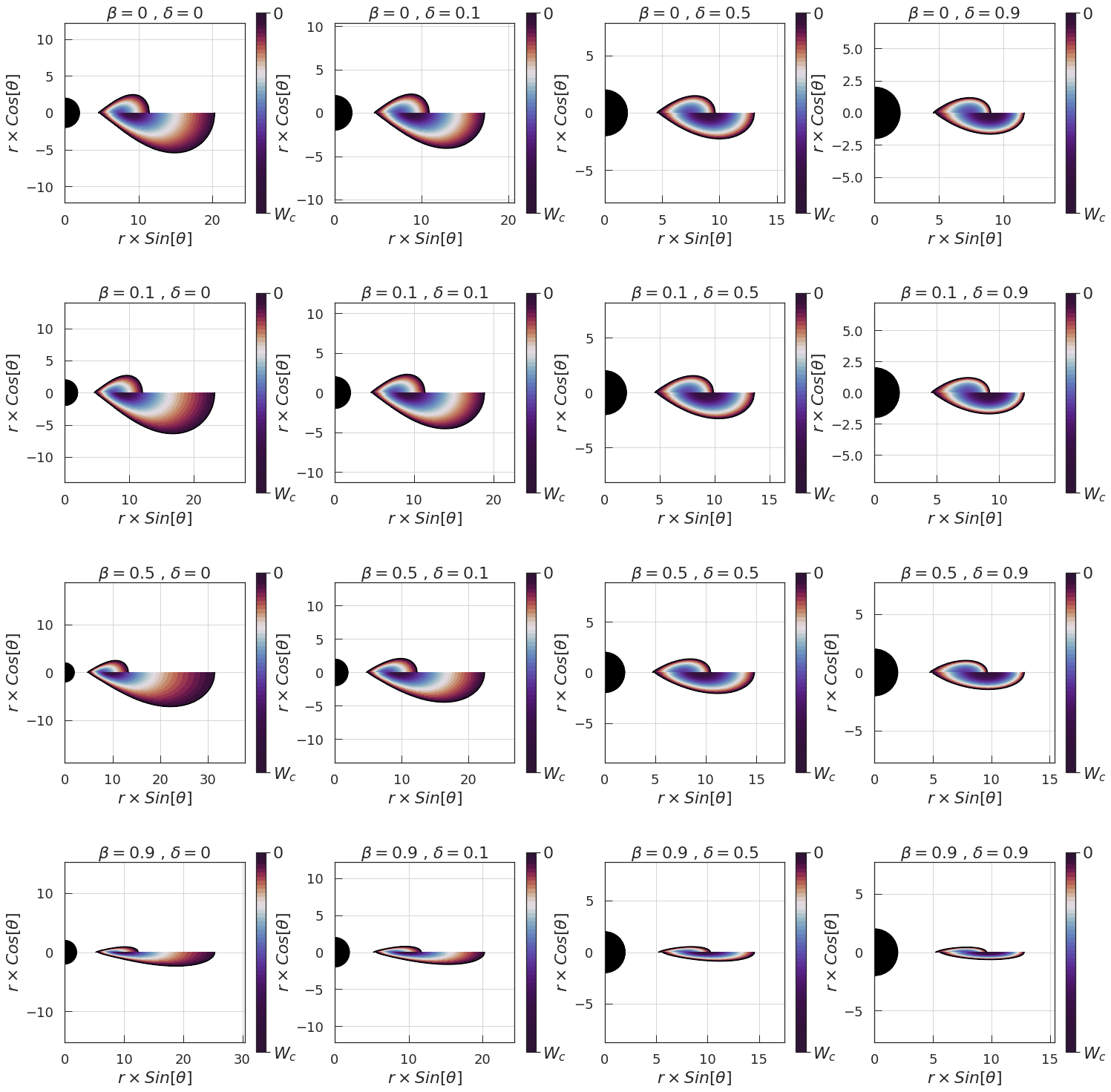}
    \caption{Map of the equipotential surfaces for various combination of $\beta$ and $\delta$. On each plot the upper part coincide to negative electromagnetic parameter $\mu=-0.01$ and the lower part to solution with positive $\mu=0.005$. All along the plot the disk orbit around a Schwarzschild Black hole.}
    \label{fig:EquiWNonConstS}
\end{figure*}

\begin{figure*}
    \centering
    \includegraphics[width=\hsize]{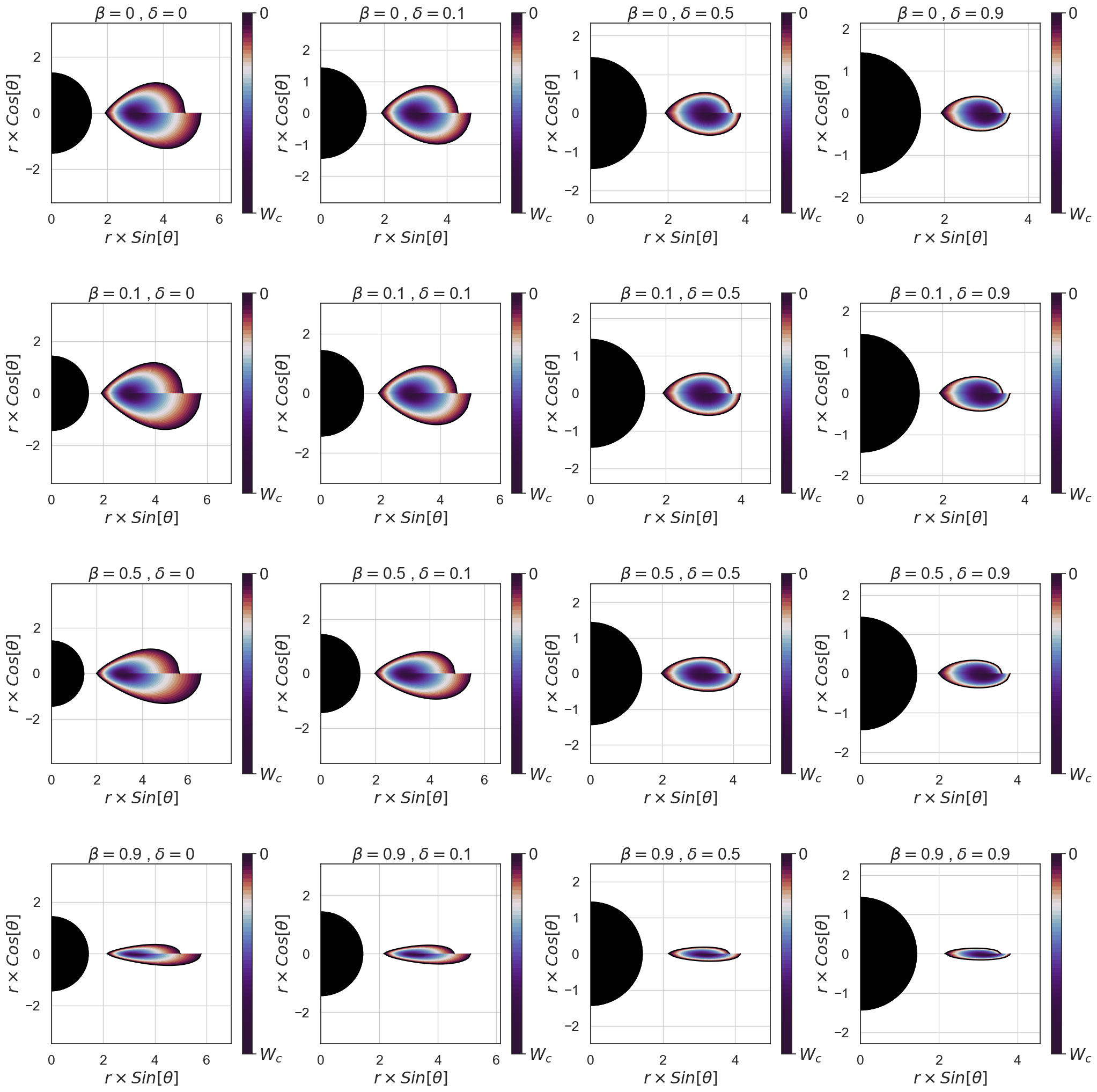}
    \caption{Map of the equipotential surfaces for various combination of $\beta$ and $\delta$. On each plot the upper part coincide to negative electromagnetic parameter $\mu=-0.01$ and the lower part to solution with positive $\mu=0.005$. All along the plot the disk orbit around a Kerr Black hole with $a=0.9$.}
    \label{fig:EquiWNonConstantK}
\end{figure*}

In this section, we explore the influence of the trigonometric angular momentum distribution on the CAD model. The Fig. \ref{fig:EquiWNonConstS} and the Fig. \ref{fig:EquiWNonConstantK} displays various combinations of $(\beta,\alpha)$ and illustrates how the equipotential surfaces behave in Schwarzschild and Kerr respectively. The upper part of each plot of both figures corresponds to negative values of the electromagnetic parameter $\mu$, while the lower part represents positive values. Throughout this configuration, the effect of the electromagnetic parameter remains consistent for both spacetime. However, we observe that increasing $\beta$ tends to enlarge the radial size of the torus. This effect is more pronounced for positive values of $\mu$. This observation aligns with the results obtained from the previous study in the equatorial plane (Section \ref{sec:EqPlaneNonConst}), where it was discussed that an increase in $\beta$ leads to a decrease in pressure gradients in the center and outer regions of the disk leading to a more extended structure in the radial direction and a thinner structure. As in the constant constant angular momentum case, dealing with a Kerr black hole affects the effect of $\mu$ in a major way. Solutions built with negative and positive $\mu$ are quite similar and looks more like a torus than a disk.

The $\delta$-parameter has a variety of effects on the equilibrium structure. As a first note, it reduces the disk's radial size by counteracting $\beta$'s effect. Additionally, it acts together with $\beta$ to reduce the vertical size of the disk by making it thinner. Furthermore, we can discuss the influence of $\delta$ on the equipotential surfaces. Increasing $\delta$ results in a more evenly distributed matter. When $\delta$ decreases, matter is concentrated in the outer part of the disk. In summary, the parameter $\delta$ plays a significant role in shaping the equilibrium structure of the accretion disk. It affects both the radial and vertical size of the disk and determines the distribution of matter across its surface. Together with $\beta$ and other parameters, $\delta$ contributes to the overall behavior and characteristics of the disk in the specific context under investigation.

\section{Pressure, Mass density and charge density profile}
\label{sec:Pressure}
\subsection{Physical limitation of the parameters}

In this section, we calculate the pressure, mass density and charge density profile resulting from the solutions built in the previous sections. To accomplish this, we must define the parameters: $n$, $\kappa$, and $B$. Initially, we adopt a polytropic index of $n=3$ to represent a relativistic fluid. The primary considerations involve determining suitable values for $B$ and $\kappa$ in a manner that ensures the validity of our fluid approximation and minimizes the self-interactions of the CAD. Choosing a weak ambient magnetic field, as the galactic magnetic field ($\tilde{B} \leq 10^{-4}$ T), is not favorable, as it would be too weak and necessitate high specific charges to maintain strong interactions. To address this, the tori would either need to be exceedingly small or highly diluted, conflicting with our fluid-based approach.

As outlined in the Introduction, an ideal scenario for achieving a nearly homogeneous magnetic field involves a black hole in a binary system paired with a magnetar, generating a magnetic field with an amplitude reaching $\tilde{B} \leq 10^{12}$ T on its surface. At distances relatively far from the magnetar, the magnetic field approximates homogeneity for latitudes near the equatorial plane. Setting the amplitude of the $B-$field to $10^{-8}$ results in $\tilde{B}=2.07 \times 10^{5}$ T in SI units.

To ensure the validity of our fluid approach, we consider the upper limit for the applicability of kinetic theory (rarefied fluids), typically up to a number density of $10^{24}$ $\rm m^{-3}$. Assuming particles of proton mass (with a specific charge $q_s \sim 10^{18}$), this imposes a mass density limit of $\rho_{\text{MHD}} \geq 10^{-3}$ kg $\rm m^{-3}$. It's important to note that our fluid is not assumed to be composed of proton plasma; rather, the specific charge profiles in our tori are several orders of magnitude lower (ranging from six to nine orders). This average specific charge represents a bulk of particles, such as a charged particle (of proton mass) surrounded by a massive numbers of neutral particles (of neutron mass). To support this scenario, the limit for applicability of our approach must be thus considered several orders higher than $\rho_{\text{MHD}}$. To achieve that, we can manipulate the parameter $\kappa$ which is associated with the density of the fluid. Opting for higher $\kappa$ values results in a more rarefied fluid. In line with our established practice in previous studies, we set $\kappa=10^5$ in this work.


The selected amplitude for $B,$ coupled with the choices of $\vert \mu \vert \geq 10^{-3}$ and $\kappa,$ yields solutions with specific charges in the range of $q_s \sim 10^7 - 10^9$, a mass density reaching $\rho_c \sim 10^{-22}-10^{-20}$ (in SI units, $\tilde{\rho_c}=10^{-1}-10$ kg $\rm m^{-3}$), and a magnetic field amplitude attaining $\tilde{B_d}= 1 - 100$ T. These values align with our initial assumptions and provide a coherent basis for our analysis.

\subsection{Analyse of the fluid properties}
\begin{figure}
\begin{tabular}{ccc}
\includegraphics[width=0.3\hsize]{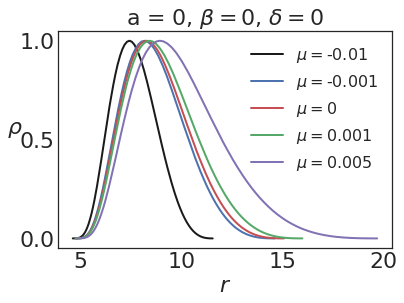}&
\includegraphics[width=0.32\hsize]{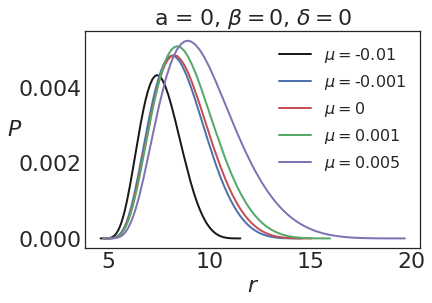}&
\includegraphics[width=0.33\hsize]{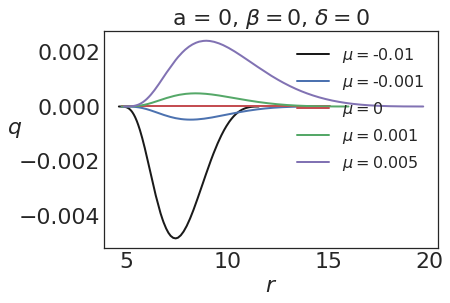}
\end{tabular}
\caption{The rest-mass density, the pressure and the charge-density equatorial profiles are shown from the left to the right for various values of $\mu$. These results present the Schwarzschild case.}
    \label{fig:fig5} 
\end{figure}

\begin{figure}
\begin{tabular}{ccc}
\includegraphics[width=0.3\hsize]{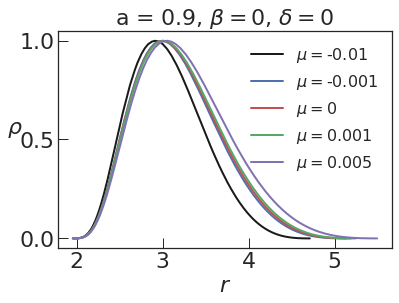}&
\includegraphics[width=0.32\hsize]{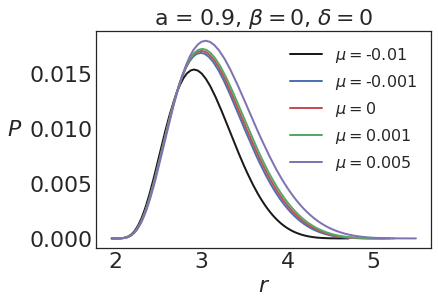}&
\includegraphics[width=0.33\hsize]{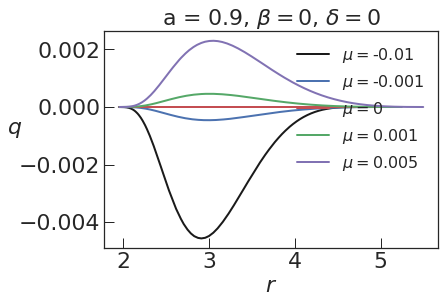}
\end{tabular}
\caption{The rest-mass density, the pressure and the charge-density equatorial profiles are shown from the left to the right for various values of $\mu$. The case of a fast rotating black hole is depicted.}
    \label{fig:fig8} 
\end{figure}

\begin{figure}
\begin{tabular}{ccc}
\includegraphics[width=0.3\hsize]{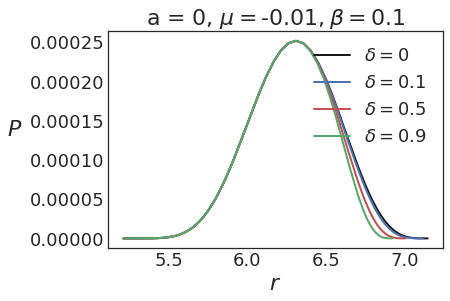}&
\includegraphics[width=0.3\hsize]{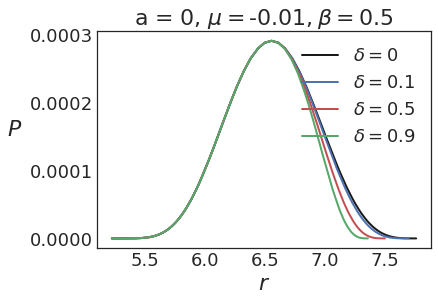}&
\includegraphics[width=0.30\hsize]{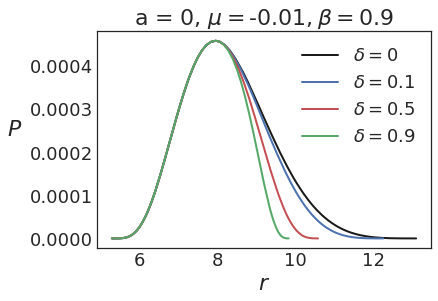}\\
\includegraphics[width=0.3\hsize]{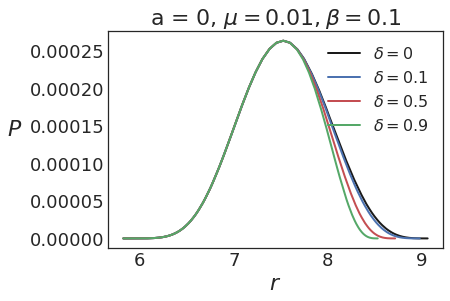}&
\includegraphics[width=0.3\hsize]{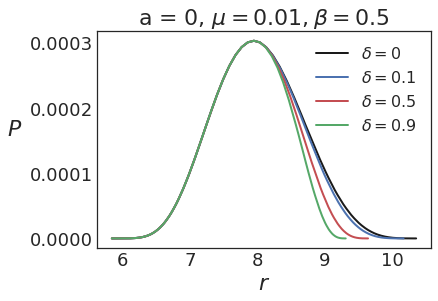}&
\includegraphics[width=0.30\hsize]{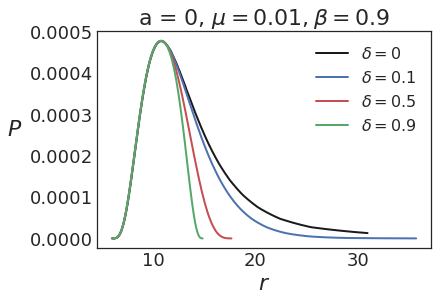}
\end{tabular}
\caption{The pressure equatorial profiles are shown from the left to the right for various values of $\beta$ and $\mu$. The Schwarzschild case is depicted.}
    \label{fig:fig6} 
\end{figure}

\begin{figure}
\begin{tabular}{ccc}
\includegraphics[width=0.3\hsize]{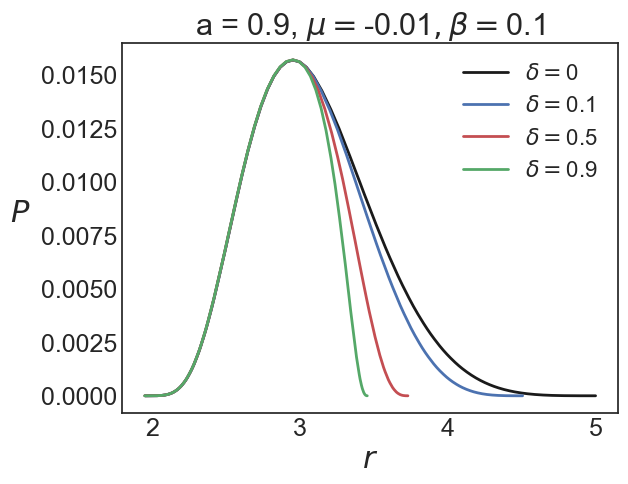}&
\includegraphics[width=0.3\hsize]{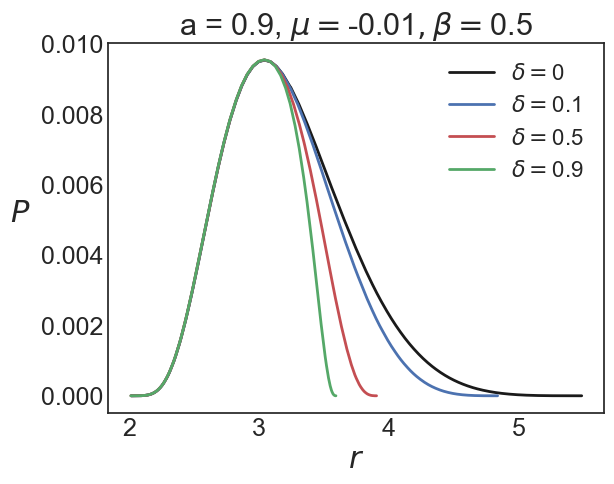}&
\includegraphics[width=0.30\hsize]{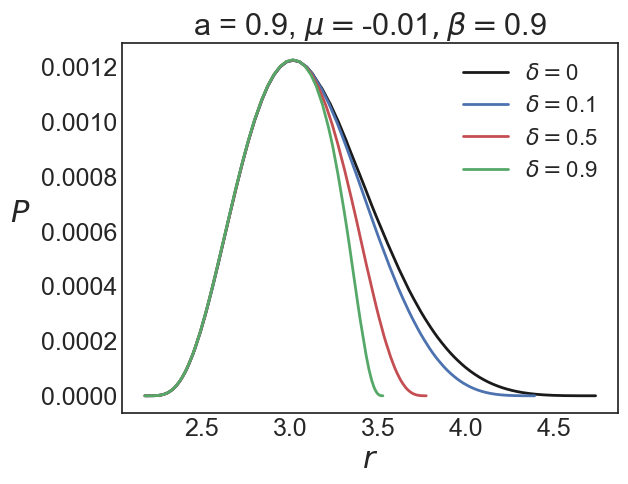}\\
\includegraphics[width=0.3\hsize]{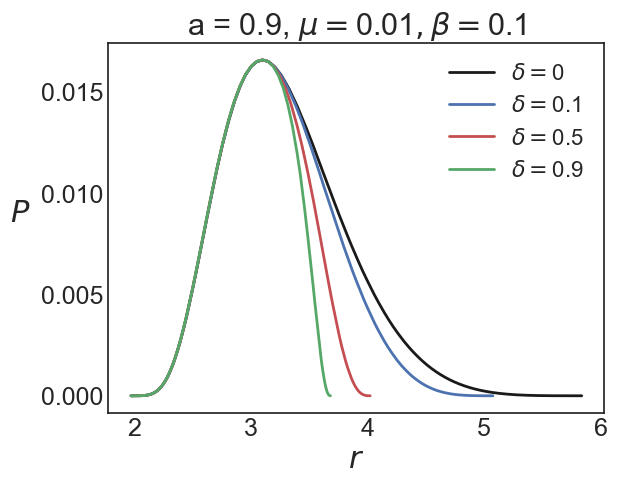}&
\includegraphics[width=0.3\hsize]{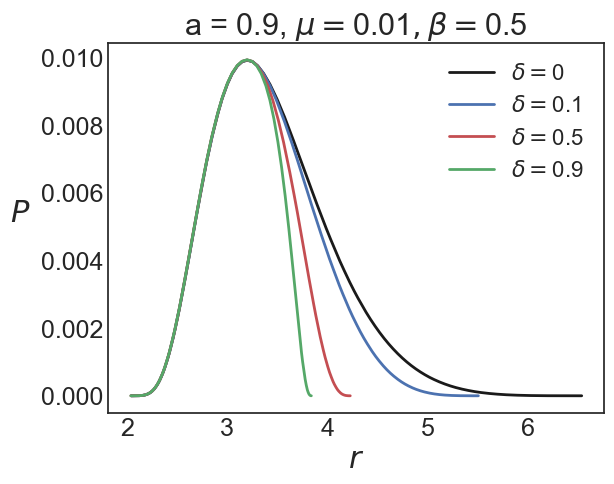}&
\includegraphics[width=0.30\hsize]{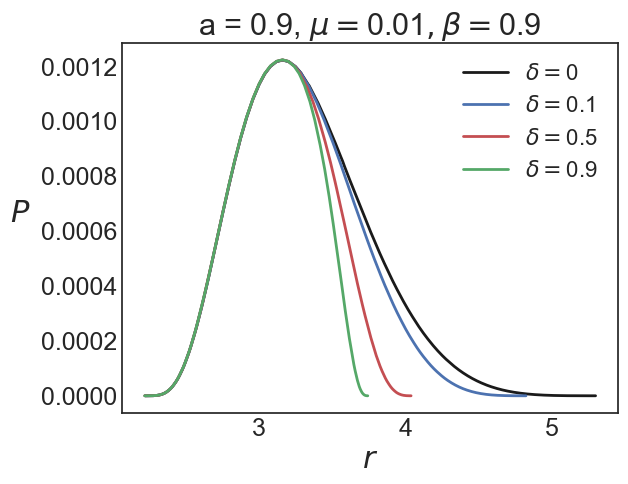}
\end{tabular}
\caption{The pressure equatorial profiles are shown from the left to the right for various values of $\beta$ and $\mu$. The Kerr case with $a=0.9$ is depicted.}
    \label{fig:fig7} 
\end{figure}

In the Figures \ref{fig:fig5}, \ref{fig:fig8},  \ref{fig:fig6} and  \ref{fig:fig7}, the density is scale with its value at the center $\rho_c$. The pressure and the charge-density are scale by $\rho_c$ to evaluate their strength relatively to the rest-mass density
In Figs. \ref{fig:fig5} and \ref{fig:fig8}, we present an analysis in the constant angular momentum case in the Schwarzschild and Kerr spacetime respectively, where we plot the rest-mass density, pressure, and charge density profiles in the equatorial plane. The rest-mass density profile reveals how mass is distributed in the equatorial plane. This profile supports the findings presented in Section \ref{sec:MapConst}. For both cases, for positive values of $\mu$, the matter is more strongly attracted to the inner part of the disk. Regarding the pressure profile, we observe that the intensity of the pressure maximum increases with an increase in the electromagnetic parameter. As we deviate from the Schwarzschild  or Kerr case, we note that negatively charged structures require less pressure to maintain equilibrium compared to positively charged disks.

The effect of non constant angular momentum is shown in the Figure \ref{fig:fig6} in the Schwarzschild case and in the Figure \ref{fig:fig7} in the Kerr spacetime with $a=0.9$.

As concluded in Fig. \ref{fig:EquiWNonConstS}, when increasing the parameter $\delta$ while keeping $\beta$ fixed, the radial size of the disk decreases. Conversely, the effect of increasing $\beta$ is the opposite, leading to an increase in the radial size of the disk. Regarding the amplitude of the maximum pressure, we find that it is influenced by $\beta$ and not by $\delta$. When $\beta$ increases, the disk requires more pressure to maintain its equilibrium. To summarize, variations in both $\beta$ and $\delta$ parameters have distinct effects on the geometric characteristics of the CAD. However, when it comes to the amplitude of the maximum pressure, $\beta$ plays a dominant role, while $\delta$ does not seem to have a significant impact. Those results are similar in Schwarzschild and in Kerr. Nevertheless, as shown when comparing the figures \ref{fig:EquiWNonConstS} and \ref{fig:EquiWNonConstantK}, the impact of the magnetic parameter is reduced when turning on the spin of the central compact object. This result could be expected considering that the magnetic field acts as a test field when the spin influence strongly the spacetime.

\section{Conclusion}

In this study, we have constructed equilibrium configurations of a charged disk orbiting a Kerr black hole, which is surrounded by an external asymptotically uniform magnetic field. The disk's angular momentum distribution was varied to investigate its impact on the system. Our findings demonstrate that these solutions share similar characteristics with previous works and the uncharged case (referred to as the "polish doughnut"). Specifically, we observed the following features:

Closed equipotential surfaces with a cusp, enabling accretion, and a pressure maximum where particles can oscillate.
The presence of a funnel, whose width depends on various parameters and allows for the possibility of a jet.
In the context of this specific background, we examined the effects of different factors, including the magnetic/charge parameter (denoted by $\mu$), the spin (referred to as $a$), and the influence of non-constant angular momentum.

In the case of constant angular momentum, we established that the radial extension of the disk is primarily governed by the spin parameter $a$. For lower values of $a$, the parameter $\mu$ also significantly impacts the size of the disk. Notably, $a$ and $\mu$ have opposing effects; increasing $a$ reduces the size of the disk, while increasing $\mu$ enlarges it. When the spin strongly deforms the disk's shape, the impact of $\mu$ becomes relatively small. Furthermore, the amplitude of the pressure maximum is affected by $\mu$: positively charged structures require more pressure to maintain equilibrium compared to negatively charged ones.

For the non-constant angular momentum case, we observed that the angular momentum profile influences the shape and size of the disk. The parameters $\beta$ and $\delta$ associated with the angular momentum have distinct effects on the disk's extension. Increasing $\delta$ reduces both the radial and vertical sizes, while a gradual increase in $\beta$ expands the radial size but contracts the vertical size significantly. The disk becomes progressively thinner at a faster rate as $\beta$ increases compared to $\alpha$. Additionally, the angular momentum profile impacts the distribution of matter within the disk, with greater values of $\delta$ leading to a less concentrated matter distribution near the inner edge. The interplay of $\mu$ with $\beta$ and $\delta$ influences the size of the disk, and the difference in radial extent between positively and negatively charged structures is more pronounced when increasing $\beta$ as opposed to $\delta$. Moreover, increasing $\delta$ brings the opposite structures closer together.

Employing a non-constant angular momentum profile inside the disk has multiple effects on the equilibrium solution's shape. The combination of such a profile with the magnetic field results in solutions that deviate from the uncharged case in the background of Kerr and Schwarzschild black holes. The impact of the charge density profile needs further exploration and will be addressed in subsequent studies. Additionally, the inclusion of more complex external magnetic fields could yield valuable insights in the future.

\section*{Acknowledgments}

AT acknowledges Shokoufe Faraji for her useful comments and feedback. Furthermore AT acknowledges the Deutsche Forschungsgemeinschaft (DFG, German Research Foundation) - funded by the Project Number $510727404$.

\bibliographystyle{ws-ijmpd}
\bibliography{sample}
\end{document}